\documentclass[aps,twocolumn,raggedbottom,prb,showpacs,nobalancelastpage,amssymb,groupedaddress]{revtex4}
\usepackage{graphicx}
\usepackage{amsmath}
\begin{document}
\title{Spectroscopy of a driven solid-state qubit coupled to a structured
environment}
\author{M.\ C.\  Goorden$^1$, M.\  Thorwart$^2$, and M.\ Grifoni$^3$}
\affiliation{$^1$Instituut-Lorentz, Universiteit Leiden, P.O. Box 9506,
2300 RA Leiden, The Netherlands\\
$^2$Institut f\"{u}r Theoretische Physik IV,
Heinrich-Heine-Universit\"{a}t D\"{u}sseldorf, 40225
D\"{u}sseldorf, Germany\\$^3$Institut f\"{u}r Theoretische Physik,
Universit\"at Regensburg, 93035 Regensburg, Germany}
\date{\today}
\begin{abstract}
We study the asymptotic dynamics of a driven spin-boson system
where the environment is formed by a broadened localized mode.
Upon exploiting an exact mapping, an equivalent formulation of the
problem in terms of a quantum two-state system (qubit) coupled to
a harmonic oscillator  which is itself Ohmically damped, is found.
We calculate the asymptotic population difference of the two
states in two complementary parameter regimes.  For weak damping
and low temperature, a perturbative Floquet-Born-Markovian master
equation  for the  qubit-oscillator system can be solved. We find
multi-photon resonances corresponding to transitions in the
coupled quantum system and calculate their line-shape
analytically. In the complementary parameter regime of strong
damping and/or high temperatures, non-perturbative real-time path
integral techniques yield analytic results for the resonance line
shape. In both regimes, we find very good agreement with exact
results obtained from a numerical real-time path-integral
approach. Finally, we show for the case of strong detuning between
qubit and oscillator  that the width of the $n$-photon resonance
scales with the $n$-th Bessel function of the driving strength in
the weak-damping regime.
\end{abstract}
 \maketitle

\section{Introduction}

Currently, we witness an impressive progress in realizing coherent
quantum dynamics of macroscopic solid state devices \cite{Nak99,Mar02,Pas03,Vio02,Chi03}. Very
recently, experimental results on the quantum dynamics of a
superconducting flux qubit coupled to a read-out SQUID have been
reported \cite{Chiorescu04}. The flux qubit consists of a
superconducting ring with three Josephson junctions and, in the
proper parameter regime, it forms a quantum mechanical macroscopic
two-state system (TSS). An external time-dependent driving force
controls the state of the TSS. A SQUID couples inductively to the
qubit and, together with an external shunt capacitance, it can be modeled 
 as a harmonic oscillator (HO). Due to the coupling of the SQUID to the
surrounding environment, the harmonic oscillator is (weakly)
damped. The state of the qubit can be inferred from the state of the SQUID. The experiment provides spectroscopic data on the
different transition frequencies of the coupled TSS-HO device.
Moreover, Rabi oscillations involving
 different pairs of quantum states of the device have been revealed, including
 the so-termed red and blue sideband transitions between energy states of the
 coupled TSS-HO system.

 Heading for a comprehensive detailed understanding, a quantitative
modeling which includes the effects of time-dependent driving,
decoherence and dissipation is required. Our description goes beyond the well-known Jaynes-Cummings model \cite{Cohbook}, by avoiding the strong rotating-wave approximation and by including a microscopic model for the environment.
A generic theoretical
model for studying the environmental effects on a driven TSS is
the driven spin-boson model \cite{Wei,Gri98} where the TSS
tunneling splitting is denoted by $\Delta$. The environment is
characterized by a spectral density $J(\omega)$. The widest used
form is that of an Ohmic spectral density, where $J(\omega)$ is
proportional to the frequency $\omega$. It mimics the effects of
an unstructured Ohmic electromagnetic environment. In the classical limit this leads to white noise and all
transitions in the system are damped equally. However, if the
environment for the qubit is formed by a quantum measuring device
which itself is damped by Ohmic fluctuations, the simple description 
as an Ohmic environment might become inappropriate. In particular, the
SQUID-detector being well described as a HO can equally well be 
considered as a (broadened) localized 
mode of the environment influencing the qubit as the 
central quantum system. In this picture, the plasma resonance at frequency
$\Omega_p$ of the SQUID gives rise to
 a non-Ohmic effective spectral density $J_{\rm eff}(\omega)$
for the qubit with a Lorentzian peak at the plasma frequency of the detector
\cite{Tian02}.

The effects of such a structured spectral density on decoherence
have been investigated in several theoretical works in various
limits. The role of the external driving being in resonance with
the symmetric TSS at zero temperature has been studied in Ref.\
\onlinecite{Thorwart00} within a Bloch-Redfield formalism being
equivalent to a perturbative approach in $J_{\rm eff}$. Smirnov's
analysis \cite{Smi03} is based on the assumption of weak
interaction between the TSS and the HO being equivalent to a
perturbative approach in $J_{\rm eff}$ as well. Moreover, a
rotating-wave approximation is used. The first assumption,
however, might become problematic if the driven TSS is in
resonance with the HO. The results presented in Refs.\
\onlinecite{Tho03,Kle03} reveal in fact, for the undriven case,
that  a perturbative approach in $J_{\rm eff}$ breaks down for strong
qubit-detector coupling, and when the qubit and detector
frequencies are comparable. Dephasing times at zero temperature
have been determined for the undriven spin-boson model with a
structured environment in Ref.\ \onlinecite{Kle04} within a
numerical flow equation method.

The interplay between the external driving and the dynamics of the coupled
TSS-HO system yields to additional
multi-photon transitions, which can be explained only by considering
the spectrum  of the coupled system. These
resonances have recently been observed experimentally \cite{Chiorescu04}.
If the time-scale of the HO does not play a role, the
multi-photon resonances occur in the driven qubit solely, which
happens when the driving frequency (or integer multiples of it)
matches the characteristic energy scales of the qubit \cite{Gri98}.
Such multiphoton resonances
 can be experimentally detected in an ac-driven flux qubit
 by measuring  the
  asymptotic occupation probabilities of the qubit, as the dc-field
  is varied  \cite{Saito04}. In these experiments the resonances
  were obtained by matching the frequency of the ac-field with the
  qubit energy levels only, where the detector energy levels did not
  play a role. These qubit resonances,
 which have also been theoretically investigated within  a
 Bloch equation formalism  in Ref.\ \onlinecite{Goo03},
 could be explained in terms  of intrinsic
 transitions in a driven spin-boson system with an unstructured environment.

In this paper, we provide a comprehensive theoretical description
of the driven spin-boson system in the presence of a structured
environment with one localized mode. Upon making use of the
equivalence of this generic model with the model of a driven TSS
coupled to an Ohmically damped HO, we first consider the
experimentally most interesting case of low temperature and
 weak damping of the HO while the coupling between the TSS
and the HO is kept arbitrary. In this regime, a
Floquet-Born-Markov master equation can be established for the
driven TSS-HO system. A restriction to the most relevant energy
states allows the analytic calculation of the asymptotic TSS
time-averaged population $P_\infty$, including the explicit shape
of the resonance peaks and dips. We furthermore consider the case
of strong damping and/or high temperature which is the
complementary parameter regime. An analytic real-time
path-integral approach within the non-interacting blip
approximation for the driven TSS with the Lorentzian-shaped
spectral density allows to analytically determine $P_\infty$ as
well. We compare the results obtained from closed analytic
expressions with those of numerically exact real-time QUAPI
calculations in both parameter  regimes and find a very good
agreement validating our analytical approaches. Finally, we
consider the weakly damped TSS with the localized mode in the
limit of large HO frequencies. Then,
the localized mode acts as a high-frequency cutoff and the usual
Ohmically damped driven TSS is recovered. For this case, we employ
an approximation valid for large driving frequencies and obtain a
simple expression for the resonance line shapes for multi-photon
transitions. Most importantly, we find that the width of the
$n$-photon resonance scales with the $n$-th ordinary Bessel
function. Parts of our results have been published in a short work
in Ref.\  \onlinecite{Goo04}.

The paper is organized as follows: In Sec.\ \ref{sec.model}, we
present the theoretical model. Then, we  treat the regime of weak
damping and low temperatures in Sec.\  \ref{sec.weakdamp}. The
complementary regime of strong damping is investigated in Sec.\ 
\ref{sec.strongdamp}. The subsequent Sec.\ \ref{sec.bessel}
contains the limit when the localized mode provides a high-frequency 
cut-off for the bath, and
Sec.\  \ref{sec.conclusio} the discussion of the results and the
conclusions. Details of the specific evaluation of rate
coefficients are presented in Appendix \ref{app.sym}. In Appendix
\ref{app.niba} an
 expansion used in the strong coupling regime is elaborated in detail.

\begin{figure}
\includegraphics[width=90mm,keepaspectratio=true,angle=0]{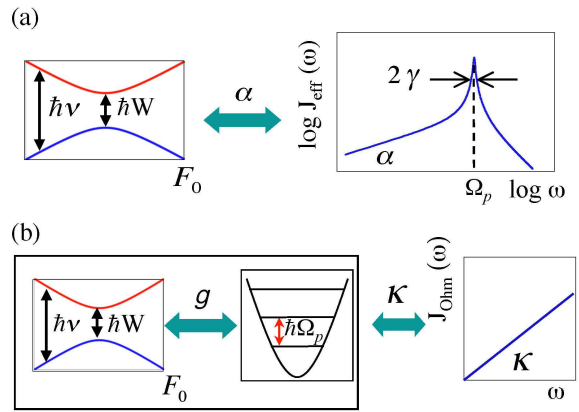}
\caption{Schematic picture of the models we use. In ($a$) the
TSS is coupled to an environment which has a peaked spectral
density $J_{\rm eff}(\omega)$. In ($b$) the system is shown as a
two-level system coupled to a harmonic oscillator which is itself
coupled to an Ohmic environment with spectral density $J_{\rm
Ohm}(\omega)$ .} \label{fig.method}
\end{figure}

\section{The driven qubit coupled to a macroscopic detector \label{sec.model}}

The driven TSS is described by the Hamiltonian

\begin{eqnarray}
H_Q(t)=-\frac{\hbar\Delta}{2} \sigma_x
-\frac{\hbar\varepsilon(t)}{2} \sigma_z\;,
\end{eqnarray}
where $\sigma_i$ are Pauli matrices, $\hbar\Delta$ is the
tunnel splitting, and  $\varepsilon(t)=\varepsilon_0 +
s\cos(\Omega t)$ describes the combined effects of a
time-dependent  driving and 
the static bias $\varepsilon_0$. In
the absence of ac-driving ($s=0$), the level splitting of
the isolated TSS is given by
\begin{equation}
\hbar\nu=\hbar\sqrt{\varepsilon_0^2+\Delta^2}\, .
\end{equation}
 The detector can be associated as part of the TSS
 environment as a localized mode.
 This gives
 the spin-boson Hamiltonian $H_{SB}$ reading
 \cite{Wei,Gri98,Thorwart00,Smi03,Tho03,Kle03,Kle04}
\begin{eqnarray}
{H_{SB}}(t)&=&H_Q(t)+\frac{1}{2}\sigma_z \hbar \sum_k \tilde{\lambda}_k
(\tilde{b}_k^{\dagger} +\tilde{b}_k) + \sum_k \hbar
\tilde{\omega}_k \tilde{b}_k^{\dagger} \tilde{b}_k.  \nonumber \\
\label{hamtot}
\end{eqnarray}
Here $\tilde{b}_k$ and $\tilde{b}_k^{\dagger}$ are annihilation
and creation operators of the $k-$th bath mode with frequency
$\tilde{\omega}_k$. The presence of the detector determines
the shape of the spectral density. Following Ref.\
\onlinecite{Tian02}, the dc-SQUID can be modeled as an effective
inductance which is shunted with an on-chip capacitance. This
gives rise to the  effective spectral density
\begin{eqnarray}
 J_{\rm eff}(\omega)& = & \sum_k \tilde{\lambda}_k^2
\delta(\omega-\tilde{\omega}_k) \nonumber \\
& = & \frac{2 \alpha \omega
\Omega_{p}^4}{(\Omega_{p}^2-\omega^2)^2+(2 \pi \kappa\omega
\Omega_{p})^2}
\label{jeff}
\end{eqnarray}
of the bath having a Lorentzian peak of width
 $\gamma=2\pi\kappa\Omega_p$ at the
characteristic detector frequency $\Omega_{p}$. It
behaves Ohmically at low frequencies
with the dimensionless coupling strength $\alpha
= \lim_{\omega \rightarrow 0} J_{\rm eff}(\omega)/2\omega$.
The qubit dynamics is described by the reduced density operator
$\rho(t)$ obtained by tracing out   the bath degrees of freedom.
The relevant observable which corresponds to the experimentally measured
switching probability of the SQUID bias current is
the population difference
$P(t):=\langle\sigma_z\rangle(t)=tr[ \rho (t) \sigma_z]$ between
the two localized states of the qubit. We focus on the asymptotic
value averaged over one period of the external driving field, i.e.,
 $P_\infty=\lim_{t\rightarrow\infty}\langle
P(t)\rangle_\Omega$.

In the following, it will become clear that it is convenient to
 exploit  the exact one-to-one
mapping \cite{Garg85} of the Hamiltonian (\ref{hamtot}) onto that of a driven TSS
coupled to a single harmonic oscillator 
 mode with frequency $\Omega_{p}$ with interaction strength $g$.
 The HO  itself interacts with a set of mutually non-interacting
 harmonic oscillators.  The corresponding total Hamiltonian is then
\begin{equation}
H_{QOB} (t) =  H_{QO} (t) + H_{OB}
\end{equation}
with
\begin{eqnarray}
H_{QO} (t)&=& H_Q(t)+ \hbar g \sigma_z (B^{\dagger}+B )  +
\hbar \Omega_{p} B^{\dagger} B\, , \nonumber \\
H_{OB}  & = &  (B^{\dagger}+B ) \sum_k \hbar \nu_k
(b_k^{\dagger} + b_k) + \sum_k \hbar \omega_k b_k^{\dagger} b_k  \, \nonumber\\
& & +  (B^{\dagger}+B )^2 \sum_k \hbar \frac{\nu_k^2}{\omega_k} \, ,
 \label{hamtlsosc}
\end{eqnarray}
where we have omitted the zero-point constant energy terms.
Here, $B$ and $B^{\dagger}$ are the annihilation and creation
operators of the localized HO mode,  while $b_k$
and $b_k^{\dagger}$ are the corresponding bath mode operators. The
spectral density of the continuous bath modes is now Ohmic with
dimensionless damping strength $\kappa$,  i.e.,
\begin{equation}
J_{\rm Ohm}(\omega) = \sum_k \nu_k^2 \delta(\omega-\omega_k) =
\kappa\omega \frac{\omega_D^2}{\omega^2+\omega_D^2} \, ,
\label{johm}
\end{equation}
where we have introduced a high-frequency Drude cut-off at
frequency $\omega_D$. If $\omega_D$ is larger than all other energy scales the particular choice of cut-off does not influence the results at long times. The relation between $g$ and $\alpha$
follows as $g=\Omega_{p} \sqrt{\alpha/(8 \kappa)}$. Fig.\
\ref{fig.method} illustrates a sketch of the two  equivalent
descriptions of the system. Fig.\  \ref{fig.method}a shows the
viewpoint where the localized mode is part of the environmental
modes, while Fig.\  \ref{fig.method}b depicts the perspective of
the localized mode being part of the ``central'' quantum system
which itself is coupled to an Ohmic environment. The equivalence
of both standpoints has first been pointed out by Garg {\em et
al.\/} \cite{Garg85} in the context of electron transfer in
chemical physics. As shown below, the first way is more
convenient for the description in terms of analytic real-time
path-integrals (Sec.\  \ref{sec.strongdamp}), while the second
viewpoint is more appropriate for the regime of weak-coupling and
for the numeric treatment with QUAPI (see below). Note that the
TSS reduced density operator $\rho(t)$ is obtained after tracing
out the degrees of freedom of the bath and of the HO. Further
progress relies on approximations which depend on the choice of
the various parameters.

\section{Weak coupling: Floquet-Born-Markov master equation\label{sec.weakdamp}}

If the coupling between the HO and the bath is weak, i.e.,
$\kappa\ll 1$ , we can choose an approach in terms of a
Born-Markov master equation in an extended Floquet description
\cite{Blu89,Kohler97,Gri98}. For a self-contained discussion, we
shortly introduce below the required formalism of the Floquet
theory. The interested reader can find more details in the review
in Ref.\ \onlinecite{Gri98}.

\subsection{Floquet formalism and Floquet-Born-Markovian master equation}
For systems with periodic driving it is convenient to use the Floquet
formalism that allows to treat periodic forces of arbitrary strength
and frequency \cite{Blu89}. It is based on the fact that the eigenstates of a
periodic Hamiltonian $H_{QO}(t)=H_{QO}(t+2\pi/\Omega)$ are of the form
\begin{eqnarray}
|\psi (t)\rangle&=&e^{-i\varepsilon_\alpha t/\hbar}|\phi_\alpha(t)\rangle \, ,\nonumber\\
|\phi_\alpha(t)\rangle&=&|\phi_\alpha(t+2\pi/\Omega)\rangle \, ,
\label{Floquetstates}
\end{eqnarray}
with the Floquet states $|\phi_{\alpha}(t)\rangle$ being periodic in time
(as is the Hamiltonian) and $\varepsilon_\alpha$ are called the Floquet or
quasi-energies. They can be obtained from the eigenvalue equation
\begin{eqnarray}
\left( H_{QO}(t)-i\hbar\frac{\partial}{\partial t} \right)|\phi_{\alpha}(t)\rangle=
\varepsilon_{\alpha}|\phi_{\alpha}(t)\rangle.
\label{floqueteq}
\end{eqnarray}
If the quasi-energy $\varepsilon_\alpha$ is an
eigenvalue with Floquet state $|\phi_\alpha(t)\rangle$, so is
$\varepsilon_\alpha+n\hbar\Omega$ with Floquet state $\exp{(in\Omega
t)}|\phi_\alpha(t)\rangle$. Both Floquet states correspond to the
same physical state.
Because of their periodicity both the Floquet states and the Hamiltonian
can be written as a Fourier series, i.e.,
\begin{eqnarray}
|\phi_\alpha(t)\rangle&=&\sum_n|\phi_{\alpha}^{(n)}
\rangle\exp{(in\Omega t)}\, ,\nonumber\\
H_{QO} (t) &=&\sum_n H_{QO}^{(n)} \exp{(in\Omega t)}.
\label{fourier}
\end{eqnarray}
Substituting these Fourier decompositions
 in the eigenvalue equation (\ref{floqueteq}) gives \cite{Shi65}

\begin{eqnarray}
\sum_k (H_{QO}^{(n-k)}+n\hbar\Omega \delta_{kn})\phi_\alpha^{(k)}=\varepsilon_\alpha
\phi_\alpha^{(n)} \, .
\end{eqnarray}
This allows us to define the Floquet Hamiltonian
${\cal H}_{QO}\equiv H_{QO}(t)-i\hbar\frac{\partial}{\partial t}$
in matrix form
with the matrix  elements
\begin{eqnarray}
\langle a n|{\cal{H}}_{QO}|bm\rangle=(H_{QO}^{(n-m)})_{ab}+n\hbar\Omega\delta_{ab}\delta_{nm}.\label{floqhamfour}
\end{eqnarray}
In the notation  $|an\rangle$, $a$ refers to a basis in which to
express  the Hamiltonian $H_{QO}(t)$, while $n$ refers to the
Fourier coefficient. The eigenvectors of ${\cal{H}}_{QO}$ are the
coefficients $\phi_\alpha^{(n)}$.

The dynamics of the system coupled to a harmonic bath is conveniently described
 by an equation of motion for the
density matrix $\rho$.
Driving effects can be captured in an elegant
way by formulating  the equation of motion in the basis of Floquet states
defined in Eq.\  (\ref{Floquetstates}).
For weak coupling to the
environment, it is sufficient to  include dissipative effects to lowest
order in $\kappa$. Within this approximation, a
Floquet-Born-Markov master equation has been established \cite{Blu89,Gri98,Kohler97}.
We average the $2\pi/\Omega$-periodic coefficients of the master equation
over one period of the driving, assuming that dissipative
effects are relevant only on timescales much larger than
$2\pi/\Omega$.
This yields equations of motion for the reduced density matrix
$\rho_{\alpha\beta}(t)=\langle\phi_{\alpha}(t)|\rho(t)|
\phi_{\beta}(t)\rangle$
of the form
\begin{eqnarray}
\dot{\rho}_{\alpha\beta}(t)
=-\frac{i}{\hbar}(\varepsilon_{\alpha}-\varepsilon_{\beta})
\rho_{\alpha\beta}(t)+
\sum_{\alpha'\beta'}L_{\alpha\beta,
\alpha'\beta'}\rho_{\alpha'\beta'}(t),
\label{rhoeq}
\end{eqnarray}
with the dissipative transition rates
\begin{eqnarray}
L_{\alpha\beta,
\alpha'\beta'}& = & \sum_n(N_{\alpha\alpha',n}
+N_{\beta\beta',n})X_{\alpha\alpha',n}
X_{\beta'\beta,-n} \nonumber \\
& &
-\delta_{\beta\beta'}\sum_{\beta'',n}N_{\beta''
\alpha',n}X_{\alpha\beta'',-n}X_{
\beta''\alpha',n}\nonumber \\
& &
-\delta_{\alpha\alpha'}\sum_{\alpha'',n}N_{
\alpha''\beta',n}X_{\beta'\alpha'',-n}X_{\alpha''
\beta,n}.
\label{rates}
\end{eqnarray}
Here, we have defined
\begin{eqnarray}
X_{\alpha\beta,n}&=&\sum_k\langle
\phi^{(k)}_{\alpha}|B+B^\dagger|\phi^{(k+n)}_{\beta}\rangle,\nonumber\\
N_{\alpha\beta,n}&=&N(\varepsilon_{\alpha}-
\varepsilon_{\beta}+n\hbar\Omega),\nonumber\\
N(\varepsilon)&=&\frac{\kappa\varepsilon}{2\hbar}\left(\coth{\left(\frac{\varepsilon}{2
k_B T}\right)}-1\right).
\end{eqnarray}
We have neglected the weak quasi-energy shifts, which are of first order in the coupling to the environment. In the sequel, we will see from a comparison with exact numerical results that this approximation is well justified.
In order to be able to solve Eq.\ (\ref{rhoeq}),
 it is necessary to determine the
Floquet quasi-energies $\varepsilon_\alpha$
and Floquet states $|\phi^{(n)}_{\alpha}\rangle$. How they can be determined
perturbatively, is shown in the following subsection.
\subsection{Van Vleck perturbation theory}
\begin{figure}
\includegraphics[width=8.5cm]{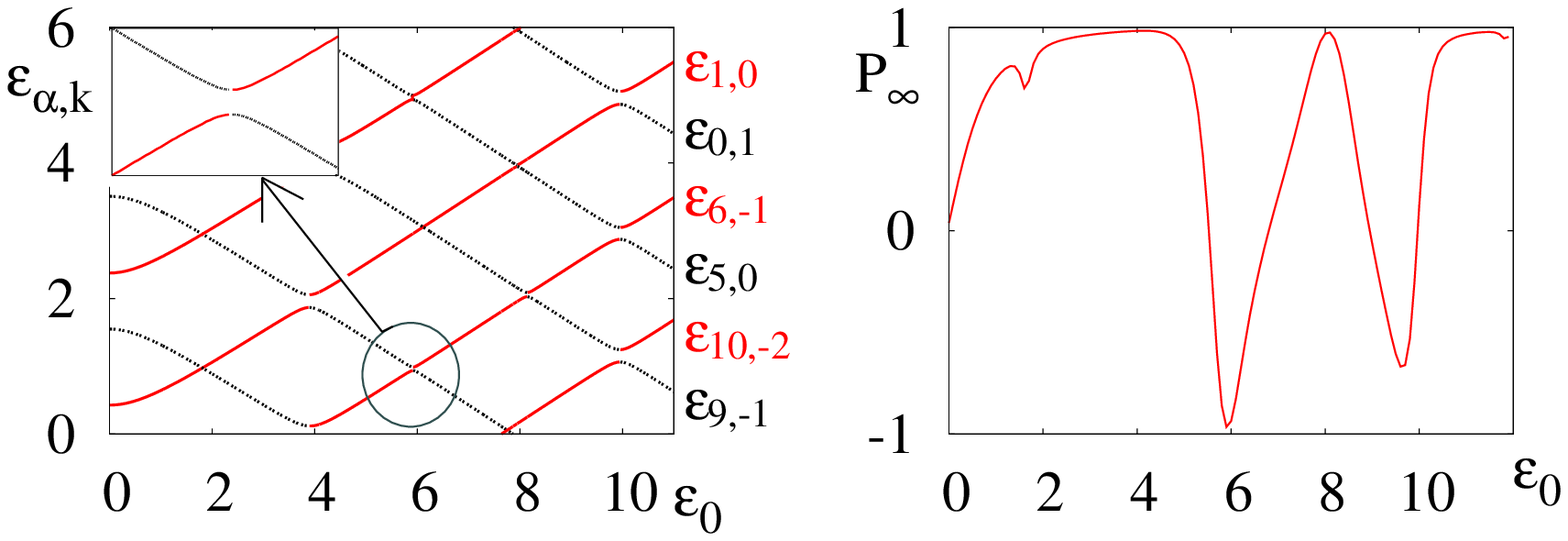}
\caption{Left: Quasi-energy spectrum $\varepsilon_{\alpha,k}$
of the
driven TSS+HO system vs dc-bias  $\varepsilon_0$ (in units of
$\Delta$). The quasi-energies are defined up to an integer multiple
of $\hbar \Omega$, i.e., $\varepsilon_{\alpha,k}=\varepsilon_{\alpha}
+k\hbar \Omega$.
Inset: Zoom of an anti-crossing. Right: $P_\infty$  exhibits resonance
dips corresponding to quasi-energy level anti-crossings.
Parameters are $\Omega=10\Delta, s=4\Delta, g=0.4\Delta, \Omega_p=4\Delta,
\kappa=0.014$ and $k_B T=0.1\hbar\Delta$.\label{spectrum}}
 \end{figure}

First, we have to specify the basis for the Floquet Hamiltonian
according to Eq.\ (\ref{floqhamfour}).   For the TSS+HO Hamiltonian
$H_{QO}$, we use the basis $|a n\rangle$ defined via  the single particle
product state $|a\rangle=|g/e\, m\rangle$ with $|g/e\rangle$
being the ground/excited state of the qubit, $|m\rangle$ the HO eigenstate,
and $n$ the corresponding Fourier index. In detail, this implies that
we can divide the Hamiltonian into a diagonal part
\begin{eqnarray}
({\cal{H}}_{{QO}})_{gmn,gmn}&=&\hbar[-\nu/2+m\Omega_p+n\Omega] \, , \nonumber\\
({\cal{H}}_{QO})_{emn,emn}&=&\hbar[\nu/2+m\Omega_p+n\Omega] \, ,
\end{eqnarray}
and an off-diagonal part
\begin{eqnarray}
({\cal{H}}_{QO})_{an,bk}&=&V_{a{n},b{k}}\, , \mbox{\hspace{2ex}
for $a\ne b, n\ne k$}\;, \label{floqdiag}
\end{eqnarray} 
which has non-zero elements. They
read
\begin{eqnarray}
V_{ g(e) ln, g(e) mn}&=&+(-)\frac{(\sqrt{m+1}\delta_{l,m+1}+\sqrt{l+1}\delta_{l+1,m})
\hbar g\varepsilon_0}{\nu} \, ,\nonumber\\
V_{g(e) ln,e(g) mn}&=&-\frac{(\sqrt{m+1}\delta_{l,m+1}+\sqrt{l+1}\delta_{l+1,m})\hbar
g\Delta}{\nu} \, , \nonumber\\
V_{g(e) mn,g(e) mk}&=&-(+)\frac{(\delta_{k,n+1}+\delta_{k+1,n})\hbar
s\varepsilon_0}{4\nu} \, , \nonumber\\
V_{g(e) mn,e(g) mk}&=&\frac{(\delta_{k,n+1}+\delta_{k+1,n})\hbar
s\Delta}{4\nu}\, .
\label{hamper}
\end{eqnarray}
In the remainder of this section we will assume that the elements of
$V$ are small compared to the diagonal elements of
${\cal{H}}_{QO}$, which is justified if the coupling $g$ between
TSS and HO and the driving amplitude $s$ are small compared to the
other energy scales, i.e., $s,g \ll \Omega,\nu,\Omega_p$. This is
the case in realistic experimental devices \cite{Chiorescu04,Tho03}. The
Fourier index $n$ ranges from $-\infty$ to $\infty$ and $m$ from
$0$ to $\infty$. The eigenvalues  of the Floquet Hamiltonian
following from  Eqs.\  (\ref{floqdiag}) and (\ref{hamper}) have to
be calculated numerically for a particular cut-off $n_{max}$ and
$m_{max}$. In Fig.\  \ref{spectrum},
 the numerically obtained quasi-energy spectrum is
shown as a function of the static bias $\varepsilon_0$
for the case $m_{\rm max}=4$ and $|n_{\rm max}|=8$.
 We find that for some values of the bias $\varepsilon_0$
 avoided crossings of the quasi-energy levels occur
  when two diagonal elements of ${\cal{H}}_{QO}$ have approximately
the same values, i.e.,  when the condition
\begin{eqnarray}
E_{a n, b m}:=({\cal{H}}_{QO})_{an,an}-({\cal{H}}_{QO})_{bm,bm}=0+O(V^2)
\end{eqnarray}
is fulfilled. It follows from Eq.\  (\ref{floqdiag}) that this happens
when at least one of the two conditions
\begin{eqnarray}
 \nu&=&n\Omega\pm m\Omega_{p}+O(V^2)\, , \nonumber \\
 n\Omega&=&m\Omega_{p}+O(V^2) \, ,
\end{eqnarray}
is fulfilled. At these avoided crossings the Floquet spectrum has quasi-degeneracies and as a consequence there are transitions between the different Floquet states. As it turns out below, this results in resonant peaks/dips in
the stationary averaged population difference $P_{\infty}$,  cf.\
Fig.\  \ref{spectrum}.

\begin{figure}
\includegraphics[width=60mm,keepaspectratio=true]{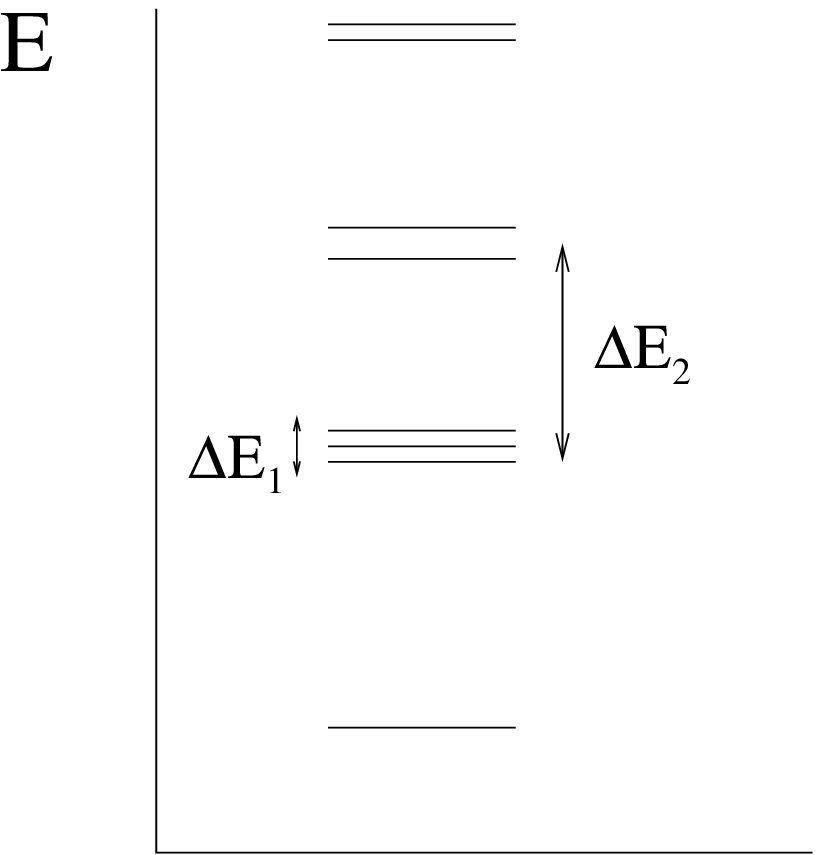}
\caption{Typical energy spectrum suited for the Van Vleck perturbation
theory: Different groups of (nearly) degenerate levels of eigenenergies are well
separated in energy (i.e.,  $\Delta E_1\ll \Delta E_2$). } \label{vleck}
\end{figure}

Since we are interested in describing the resonance line shape for $P_\infty$,
we have to determine the quasi-energies and Floquet states around
a resonance, i.e., around an avoided crossing. For this, we use
 an approach which is perturbative in $V$. The
unperturbed Hamiltonian is diagonal and, close to an avoided
crossing, (nearly) degenerate. An appropriate perturbative method
is the Van Vleck perturbation theory \cite{Cohbook,Sha80} suitable
for Hamiltonians for which the unperturbed spectrum has groups of
(nearly) degenerate eigenvalues, well separated in energy space.
An example of such a spectrum is shown in Fig.\ \ref{vleck}.
This method defines a unitary transformation which transforms the
Hamiltonian into an effective block-diagonal one. The effective
Hamiltonian then has the same eigenvalues as the original Hamiltonian, with the
quasi-degenerate eigenvalues in one common block.

The effective Hamiltonian can be written as
\begin{eqnarray}
{\cal{H}}_{\rm{eff}}=e^{iS}{\cal{H}}_{QO}e^{-iS}.
\label{gentransform}
\end{eqnarray}
In Ref.\  \onlinecite{Cohbook} it is shown how to
obtain $S$ systematically for every order in the perturbation. The small parameter is $V/\Delta E_2$ (see Fig.\ \ref{vleck}). Eigenvalues within one block can be arbitrarily close. This means that we can also use the expansion at resonance. We
derive the expressions up to the second order in the perturbation.
Two different cases are relevant: For the case when
$({\cal{H}}_{QO})_{an,an}$ and  $({\cal{H}}_{QO})_{bm,bm}$  are not nearly
degenerate we find
\begin{eqnarray}
iS_{a n,b m}^{(1)}& = &\frac{V_{a n,b m}}{E_{a n,b m}}\, , \nonumber \\
iS_{a n,b m}^{(2)}& = &
\sum_{c, k}\frac{V_{an,c k}V_{c k,b m}}{2E_{b m,a
n}}\left(\frac{1}{E_{c k,a n}}+\frac{1}{E_{c k,b m}}\right) \, , \nonumber \\
\label{transform}
\end{eqnarray}
where the superscript indicates the order of perturbation theory.
For the second case when $({\cal{H}}_{QO})_{an,an}$ and
 $({\cal{H}}_{QO})_{bm,bm}$ are nearly degenerate, we find
$iS_{a n,b m}^{(1)}=iS_{a n,b m}^{(2)}=0$.

In turn, the matrix elements of the $n$-th order term ${\cal{H}}^{(n)}_{\rm{eff}}$
of the effective Hamiltonian can be calculated, again for both cases.
When $({\cal{H}}_{QO})_{an,an}$ and $({\cal{H}}_{QO})_{bm,bm}$ are not nearly degenerate, we
find
$({\cal{H}}_{\rm{eff}})_{a n,b m}^{(1)}=({\cal{H}}_{\rm{eff}})_{a n,b m}^{(2)}
=0$. For the second case, when $({\cal{H}}_{QO})_{an,an}$ and $({\cal{H}}_{QO})_{bm,bm}$
are nearly degenerate, one finds
\begin{eqnarray}
({\cal{H}}_{\rm{eff}})_{a n,b m}^{(1)}&=&V_{a n,b m}\, , \nonumber\\
({\cal{H}}_{\rm{eff}})_{a n,b m}^{(2)}&=& \frac{1}{2}\sum_{c, k}{V_{a
n,c k}V_{c k,b m}}\left(\frac{1}{E_{a n,c k}}
+\frac{1}{E_{b m,c k}}\right)\, .\nonumber\\
\label{newenergy}
\end{eqnarray}
%

%
%
Since the effective block-diagonal Hamiltonian consequently only
has $2 \times 2$ blocks, it is easy to  diagonalize it. To obtain the
eigenvectors of the original Floquet Hamiltonian (see Eqs.\
(\ref{floqdiag},\ref{hamper})),  the inverse of the transformation
defined in Eq.\  (\ref{gentransform}) has to be performed on the
eigenvectors. There is an infinite number of quasi-energy levels
and Floquet states. However, because the eigenvalues
$\varepsilon_\alpha$ and $\varepsilon_\alpha+n\hbar\Omega$
represent the same physical state, only one of them has to be
considered. Still there is an infinite number of levels because
the Hilbert space of the HO Hamiltonian is infinite dimensional.
Nevertheless, for practical calculations, only the relevant HO
levels have to be taken into account. When there is a resonance
between the states $| e/g,0, n\rangle$ and $|e/g,l, n+k\rangle$,
then at least the first $l$ levels of the HO play a role. Higher
levels can be omitted if one is interested in low temperatures,
which is commonly the case, since for low temperatures their
occupation number will be very small.
\subsection{Line shape of the  resonant peak/dip}
To obtain the line shape of the resonant peak/dip in $P_\infty$,
we have to determine the stationary solution of  Eq.\
(\ref{rhoeq}).  Depending on the number $n_{max}$ of Floquet
states taken into consideration, this might be considerably
difficult. One facilitation might arise due to symmetries, i.e.,
elements of $L_{\alpha\beta,\alpha'\beta'}$ being of the same
size. Another possibility appropriate at low temperatures might be
to neglect some of the dissipative transition rates. Moreover, a
further possible approximation can be applied for
$\rho_{\alpha\beta}(\infty)$, if $\varepsilon_\alpha$ and
$\varepsilon_\beta$ are not nearly-degenerate eigenvalues. In that
case $\varepsilon_\alpha-\varepsilon_\beta$ in Eq.\  (\ref{rhoeq})
is much larger than the coefficients
$L_{\alpha\beta\alpha'\beta'}$, since the coupling to the Ohmic
environment is assumed to be weak. This, in turn, allows
 to make the partial secular approximation by setting
$\rho_{\alpha\beta}(\infty)=0$. After the reduced density matrix
in the Floquet basis is known, it is straightforward to calculate
$P_{\infty}$.

\subsection{Example: The first blue sideband}

As an example we will derive an analytical expression for the resonant
dip at $\nu\approx\Omega-\Omega_{p}$ which is called the first blue sideband.
For this case, the matrix elements $({\cal{H}}_{QO})_{g0\,n+1,g0\,n+1}$ and
 $({\cal{H}}_{QO})_{e1n,e1n}$ are nearly degenerate, i.e.,
\begin{equation}
 -\nu/2+(n+1)
 \Omega\approx \nu/2+\Omega_p+n\Omega \, .
 \end{equation}
  As the resonance occurs
 between two states which differ only by one oscillator quantum, we
 only take into account one excited level of the oscillator. We expect that this is
 a reasonable approximation for not too strong driving and low temperatures.
 The validity of this approximation will be checked against numerically exact
 results in the end. The elements of the transformation matrix $S$ follow as
$S_{g0\,n+1,e1n}=S_{e1n,g0\,n+1}=0$, while the remaining
elements can be calculated straightforwardly using Eq.\  (\ref{transform}) and
they will not be given here  explicitly.

Since we include one HO excited energy level, we have four physically
different eigenstates. Hence, we can express ${\cal{H}}_{QO}$ in the basis
$\left\{|e,1,-1\rangle,|g,0,0\rangle,|g,1,0\rangle,|e,0,0\rangle\right\}$.
Performing the transformation defined in Eq.\  (\ref{gentransform}),
 we obtain the effective Hamiltonian in this basis as %
\begin{widetext}
\begin{eqnarray}
H_{\rm eff} & = &
\hbar\left(\begin{array}{cccc}
\frac{\nu}{2}+\Omega_p-\Omega+ W_1&-\Delta_1&0&0\\
  -\Delta_1&-\frac{\nu}{2}+ W_2&0&0\\
  0&0&-\frac{\nu}{2}+\Omega_p+ W_3&0\\
  0&0&0& \frac{\nu}{2}+ W_4
\end{array}
\right).
\label{floqmat}
\end{eqnarray}
\end{widetext}
The matrix elements are calculated using Eq.\  (\ref{newenergy}). They read
\begin{eqnarray}
\Delta_1&=&\frac{\Delta\varepsilon_0gs\left[\Omega^2+\Omega_p^2+\nu(-\Omega+\Omega_p)
\right]}{4\nu(\Omega-\nu)\Omega\, \Omega_p(\nu+\Omega_p)}\, , \nonumber \\
W_1&=&-W_2=\frac{\varepsilon_0^2g^2}{\nu^2\Omega_p}+\frac{\Delta^2g^2}{\nu^2(\nu+\Omega_p)}
+\frac{\Delta^2s^2}{8\nu(\nu^2-\Omega^2)}\, ,\nonumber\\
 W_3&=& -W_4=\frac{\varepsilon_0^2g^2}{\nu^2\Omega_p}-\frac{\Delta^2g^2}{\nu^2(\nu-\Omega_p)}
-\frac{\Delta^2s^2}{8\nu(\nu^2-\Omega^2)} \, .\nonumber\\
\end{eqnarray}
The eigenvalues of the  Hamiltonian (\ref{floqmat})
are the relevant quasi-energies, and they are readily obtained by
diagonalization as
\begin{eqnarray}
\frac{\varepsilon_{1/2}}{\hbar} &=&-\frac{\nu}{2}+\frac{\delta}{2}
\,\left( 1\mp \sqrt{1+\frac{\Delta_1^2}{\delta^2}}\right)-
W_1\, ,\nonumber\\
\frac{\varepsilon_{3}}{\hbar}&=&-\frac{\nu}{2}+\Omega_{p}+ W_3\, ,  \nonumber\\
\frac{\varepsilon_{4}}{\hbar}&=&\frac{\nu}{2}+  W_4 \, .
\end{eqnarray}
From these formulas it follows that
$\delta=\nu-\Omega+\Omega_{p}+2 W_1$ is a measure of how far the
system is off resonance. For $\delta=0$, the quasi-energies
$\varepsilon_1$ and $\varepsilon_2$ show an avoided crossing of
size $\hbar\Delta_1$. Note that Eq.\ (\ref{rhoeq})
 implies that $\Delta_1$ is the Rabi frequency at the blue sideband.

The eigenvectors, which are the Floquet states,
of the $4\times4$ effective block-diagonal matrix in
Eq.\  (\ref{floqmat}) are easily obtained by
performing  the corresponding  inverse
transformation. We find,  with $\tan{\theta}=2|\Delta_1|/\delta$, the
eigenstates
\begin{eqnarray}
|\phi_{1}\rangle&=&e^{-iS}[\sin{(\theta/2)}e^{-i\Omega t}|e,1\rangle+
\cos{(\theta/2)}|g,0\rangle]\, ,\nonumber\\
|\phi_{2}\rangle&=&e^{-iS}[\cos{(\theta/2)}e^{-i\Omega t}|e,1\rangle-
\sin{(\theta/2)}|g,0\rangle]\,  ,\nonumber\\
|\phi_{3/4}\rangle&=&e^{-iS}|g/e,1/0\rangle \, . \label{Floquet}
\end{eqnarray}
We have used the inverse transformation of Eq.\ (\ref{fourier}) to illustrate
the time-dependence explicitly.
Next, we calculate the rates given in Eq.\ (\ref{rates}) up to second order in $V$.

The quasi-energies $\varepsilon_1$ and $\varepsilon_2$ are
quasi-degenerate. To be definite, we assume a {\em partial secular
approximation}: We set almost all off-diagonal elements of $\rho$
to zero but keep $\rho_{12}(\infty)$ and
$\rho_{21}(\infty)=\rho_{12}^*(\infty)$ different from zero.
This allows to simplify the master equation (\ref{rhoeq}).
The stationary solutions are determined by the conditions
\begin{eqnarray}
0&=&\sum_{\beta}L_{\alpha\alpha,
\beta\beta}\rho_{\beta\beta}(\infty)+(L_{\alpha\alpha,12}+L_{\alpha\alpha,21})
{\rm Re}[\rho_{12}(\infty)],
\nonumber\\
0&=&-\frac{i}{\hbar}(\varepsilon_{1}-\varepsilon_{2})\rho_{12}(\infty)+
\sum_{\alpha}L_{12,\alpha\alpha}\rho_{\alpha\alpha}(\infty)\nonumber\\
& &+ L_{12,12}\rho_{12}
(\infty)+L_{12,21}\rho_{12}^*(\infty).
\label{rhoeq2}
\end{eqnarray}
It is most convenient to use the symmetry properties of the corresponding rates
which are specified for this particular example in the Eqs.\
(\ref{symmetries}) in Appendix \ref{app.sym}. In turn, there are eight
 independent rates associated to all possible transitions. They are
 explicitly given in the Eqs.\  (\ref{allrates}).

First we consider the rates exactly at resonance $\delta=0$. Since then
$\sin^2({\theta/2})=\cos^2({\theta/2})=1/2$,  all rates contain a
term which is of zeroth order in $V$. If we neglect the small second order
terms, we find
\begin{eqnarray}
L_{22,44}&=&L_{11,44}=L_{33,22}=L_{33,11}=-L_{33,21} \nonumber\\
& =& L_{21,44}=N(\hbar\Omega_p)\;,\nonumber\\
L_{22,33}&=&L_{11,33}=L_{44,22}=L_{44,11}=L_{44,21}\nonumber\\
& = & -L_{21,33}=N(-\hbar\Omega_p)\;,\nonumber\\
L_{11,21}&=&L_{22,21}=L_{21,22}=L_{21,11}\nonumber\\
& = & \frac{1}{2}[N(\hbar\Omega_p)-N(-\hbar\Omega_p)]\;,\nonumber\\
L_{12,12}&=&-N(-\hbar\Omega_p)-N(\hbar\Omega_p)\;.
\label{ratesres}
\end{eqnarray}
Solving Eq.\ (\ref{rhoeq2}) together with (\ref{ratesres}) finally yields
\begin{eqnarray}
\rho_{11}(\infty)&=&\rho_{22}(\infty)=\frac{N(-\hbar\Omega_p)N(\hbar\Omega_p)}
{[N(-\hbar\Omega_p)+N(\hbar\Omega_p)]^2}\;,\nonumber\\
\rho_{33}(\infty)&=&\frac{N(\hbar\Omega_p)^2}{[N(-\hbar\Omega_p)
+N(\hbar\Omega_p)]^2}\;,\nonumber\\
\rho_{44}(\infty)&=&\frac{N(-\hbar\Omega_p)^2}{[N(-\hbar\Omega_p)
+N(\hbar\Omega_p)]^2}\;,\nonumber\\
\rho_{12}(\infty)&=&0\;.\nonumber\\
\end{eqnarray}
Eventually, this gives the  simple result at resonance, $\delta=0$,
\begin{equation}
P_{\infty}=-\frac{\varepsilon_0}{\nu}\tanh\left(\frac{\hbar\Omega_p}{2
k_B T}\right)+O\left(V^2\right)\ , \label{pinffloquet}
\end{equation}
which implies a complete inversion of population at low temperatures! We will discuss the physics of this in section \ref{discussion}. Note that
no further assumption on the temperature was made while deriving this formula.

\begin{figure}
\includegraphics[width=9cm]{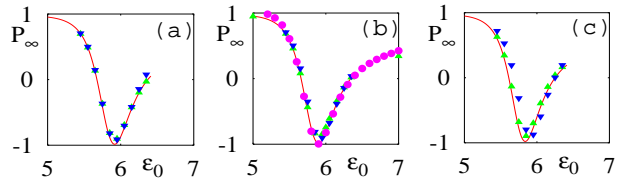}
\caption{$P_\infty$ vs $\varepsilon_0$ (in units of $\Delta$)
around the peak at $\nu=\Omega-\Omega_p$.
 The solid lines are the analytical prediction (\ref{analytical}) for
 $(a)$ $g=0.05\Delta$, $(b)$ $g=0.2\Delta$, $(c)$ $g=0.4\Delta$.
 The triangles are the results of a Floquet-Bloch-Redfield simulation, cf.\
 Eq.\  (\ref{rhoeq}),
 with one (upward triangles) and two (downward triangles) HO levels taken into account.
 The circles in ($b$) are the results from a QUAPI simulation with six HO levels
 (see text).
 We choose $s=2\Delta$,
 $\Omega=10\Delta$, $\kappa=0.014$, $k_BT=0.1\hbar\Delta$. } \label{floquet}
\end{figure}

Next we will derive an expression for the peak shape around the resonance.
For this, we assume low temperatures,
i.e., $k_B T / \hbar \ll \Omega_p, \Omega, \nu$. This allows us to
set $N(\hbar\Omega_p)=N(\hbar\Omega)=N(\hbar\nu)=0$.
Far enough away from resonance, it is appropriate to
assume that
$\rho_{12}(\infty) \approx \rho_{21}(\infty)\approx 0$,
and $\sin{(\theta/2)}\approx \theta/2$. Thus, it follows
from Eq.\  (\ref{symmetries}) that
there are only four independent rates in this case, namely
$L_{44,22},L_{22,44},L_{44,11}$ and
$L_{11,44}$. Within our approximations, we find that $L_{22,44}=O(V^3)$.
So only three rates are relevant which read
\begin{eqnarray}
L_{44,22}&=&2\cos^2(\theta/2)\approx 2\, ,\nonumber\\
L_{11,44}&=&2L_{\rm q}(\varepsilon_{1,4,0})\cos^2(\theta/2)\approx
\frac{8\Delta^2g^2\Omega_p^2}{(\nu^3-\nu\Omega_p^2)^2}\;,\nonumber\\
L_{44,11}&=&2\sin^2(\theta/2)\approx\theta^2/2\approx 2\Delta_1^2/\delta^2 \, ,
\label{dominantrates}
\end{eqnarray}
where the  quantity $L_{\rm q}$ is given in Appendix
\ref{app.sym}. Note that $L_{44,22}\gg L_{44,11},L_{11,44}$. In this limit we find for the asymptotic density matrix elements
\begin{eqnarray}
\rho_{11}(\infty)&=&\frac{L_{11,44}}{L_{11,44}+L_{44,11}}\, ,\nonumber\\
\rho_{22}(\infty)&=&\rho_{33}(\infty)=0\, ,\nonumber\\
\rho_{44}(\infty)&=&1-\rho_{11}(\infty)\,,
\end{eqnarray}
which gives the central result
\begin{eqnarray}
P_{\infty}&=&\frac{\varepsilon_0}{\nu}\frac{L_{11,44}
-L_{44,11}}{L_{11,44}+L_{44,11}}+O\left(V^2\right) \nonumber \\
&\simeq & \frac{\varepsilon_0}{\nu}\left(
1 - \frac{2\Delta_1^2 \nu^2 (\nu^2-\Omega_p^2)^2}
{\Delta_1^2 \nu^2 (\nu^2-\Omega_p^2)^2 + 4 \Delta^2 g^2 \Omega_p^2\delta^2}
\right)\, . \nonumber \\
\label{analytical}
\end{eqnarray}
A comparison between the result of this formula and different
numerical results, including those of an exact numerical ab-initio
real-time QUAPI calculation \cite{QUAPI,Tho98,Tho00}, is shown in
Fig.\  \ref{floquet}. For the QUAPI-simulations, we have used the
optimized parameters \cite{Tho00} $\Delta t=0.23/\Delta, M=12$ and
$K=1$. Moreover, we have applied an exponential cut-off for the
Ohmic bath with a cut-off frequency $\omega_c=10 \Delta$ (since we
are considering long-time stationary results, the explicit shape
of the cut-off is irrelevant). Note that the picture  of the
TSS+HO being the central quantum system which is coupled to an
Ohmic environment is particularly suited for QUAPI since the
coherent dynamics of the central quantum system is treated
exactly. A very good agreement, even near resonance,  is found
among all the used numerical schemes.

\subsection{Results and discussion}\label{discussion}

To get a qualitative understanding of the results it is important
to realize how the rates $L_{\alpha\beta,\alpha'\beta'}$ are formed.
All Floquet states are superpositions of the four unperturbed states
$|g,0\rangle$, $|g,1\rangle$, $|e,0\rangle$, $|e,1\rangle$. The rates
$L_{\rm osc}$, $L_{\rm q}$, $L_{\rm q,osc}$ defined in Eq.\ (\ref{transitions})
 describe
the time scales of the transitions between the different
unperturbed states. In this discussion these will be called the
basic transition rates. If we want to calculate the rates
$L_{\alpha\alpha',\beta\beta'}$ which describe transitions between
different Floquet states, we have to multiply the basic rates by
the square of the amplitudes in the superpositions. For example
the Floquet state $|\phi_1\rangle$ consists of an unperturbed
state $|e,1\rangle$ with amplitude $\sin(\theta/2)$, and
$|g,0\rangle$ with amplitude $\cos(\theta/2)$. The rate
$L_{44,11}$ describing the dissipative transition from
$|\phi_1\rangle$ to $|\phi_4\rangle$ has a term which is the basic
transition rate from $|e,1\rangle$ to $|e,0\rangle$ ($L_{\rm
osc}$) multiplied by $\sin^2(\theta/2)$, and a term which is the
product of the basic transition rate from $|g,0\rangle$ to
$|e,0\rangle$ ($L_{\rm q}$) and $\cos^2(\theta/2)$. In this
qualitative explanation we can neglect the amplitudes of the other
states in $|\phi_1\rangle$ and $|\phi_4\rangle$ which are
$O(V^2)$.

Now we consider the rates at resonance. Here, the amplitudes
$\sin(\theta/2)$ and $\cos(\theta/2)$ are equal and we only have to
consider the largest basic rate which is $L_{\rm osc}$. The dominant
transition from $|\phi_1\rangle$ to $|\phi_4\rangle$ is via the basic
transition from $|e,1\rangle$ to $|e,0\rangle$ and it is fast (both
amplitude in the superposition and rate $L_{\rm osc}$ are of order one).
The same holds for the transitions between $|\phi_2\rangle$ to $|\phi_4\rangle$.
There will also be a transition between $|\phi_3\rangle$ and
$|\phi_1\rangle$, $|\phi_2\rangle$ via the basic transition between
$|g,1\rangle$ and $|g,0\rangle$. Since only the decay of the  oscillator plays
a role, the stationary state is a thermally equilibrated
mixture between $|e/g,1\rangle$
and $|e/g,0\rangle$ which is described by Eq.\ (\ref{pinffloquet}).

Away from resonance, the amplitude $\sin(\theta/2)$ becomes small
($\sim \Delta_1$) implying that other basic transitions start
to play a role. The transition from $|\phi_2\rangle$ to $|\phi_4\rangle$
is still dominated by the basic transition from $|e,1\rangle$ to
$|e,0\rangle$ with large amplitude $\cos(\theta/2)$. Within our assumption of
low temperatures used  to derive Eq.\  (\ref{analytical}),
the state $|\phi_2\rangle$ is weakly populated after
long times. The same holds for $|\phi_3\rangle$. Between
$|\phi_1\rangle$ and $|\phi_4\rangle$ two basic transitions are
important. The first is a transition from $|e,1\rangle$ to $|e,0\rangle$
with large rate (i.e., fast)
but low amplitude ($\sin(\theta/2)$) (i.e., rare). For low
temperatures, this rate describes transitions from $|\phi_1\rangle$
to $|\phi_4\rangle$ and it dominates $L_{44,11}$ in Eq.\ (\ref{dominantrates}).
The other basic transition occurs via the unperturbed states
$|e,0\rangle$ and $|g,0\rangle$. The basic rate $L_{\rm q}$ is small, but
it has a large amplitude ($\cos(\theta/2)$). For low temperatures
this mechanism causes transitions from $|\phi_4\rangle$ to $|\phi_1\rangle$
and it is the dominant part of $L_{11,44}$. The stationary state is
described by the ratio of the rates as described by Eq.\ (\ref{analytical}).
Note that since both rates scale with $g^2$ the final result does not depend on
$g$, for small $g$. For other resonances we
 find a different eigenvalue splitting and the peak shape will
depend on $g$.

We will now try to give a more physical insight in the nature of this resonance. First consider the case when $T=0$ and we are exactly on resonance. The driving induces transitions from $|g,0\rangle$ to $|e,1\rangle$ while the direct coupling of the HO to the environment will cause a fast decay of the population from $|e,1\rangle$ to $|e,0\rangle$. This transition from $|g,0\rangle$ to $|e,0\rangle$ via driving and decay has to compete with the decay from $|e,0\rangle$ to $|g,0\rangle$, but the last process is much slower because the TSS is not directly coupled to the environment. So all the population is in $|e,0\rangle$ and there is a complete inversion of population. For $T\ne 0$ there will be a thermal equilibrium between ground and excited state of the oscillator.

When the system is not exactly at resonance the driving induced transitions are much slower $\propto g^2$ and the decay of the oscillator is still fast. This means that the transition from $|g,0\rangle$ to $|e,0\rangle$ is slower than at resonance. The time associated with the decay from $|e,0\rangle$ to $|g,0\rangle$ is also $\propto g^2$ and the ratio of the time scales of the two processes gives the ratio of the populations of $|e,0\rangle$ and $|g,0\rangle$ (at $T=0$, for higher $T$ the states $|g/e,1\rangle$ are also populated). This ratio is independent of $g$ and so is $P_\infty$.

A similar analysis can be performed for the first red sideband at
$\nu=\Omega+\Omega_p$. At resonance it yields
$P_\infty=\frac{\varepsilon_0}{\nu}\tanh({\frac{\hbar\Omega_p}{2k_B
T}})+O(V^2)$, which is very close to thermal equilibrium for low
$T$.

For the resonance at $\nu=\Omega_p$,
 only the oscillator is excited.
After having traced it out,  we expect just thermal equilibrium given by
$P_\infty=\frac{\varepsilon_0}{\nu}\tanh({\frac{\hbar\nu}{2 k_B
T}})$.

\section{Strong coupling: NIBA \label{sec.strongdamp}}

\begin{figure}
\includegraphics[width=5cm,angle=270]{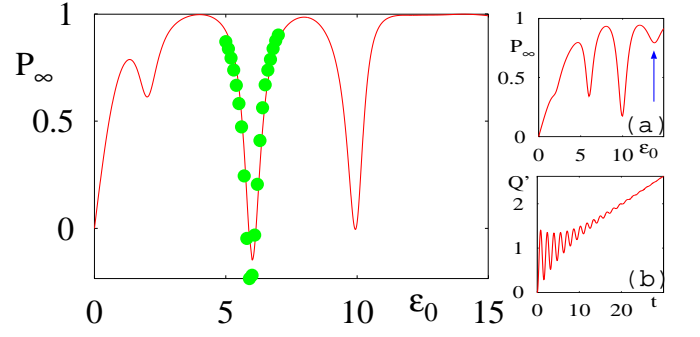}
\caption{ $P_\infty$ vs $\varepsilon_0$ (in units of $\Delta$).
The solid line is the NIBA prediction, while the circles are from
a QUAPI simulation with 6 HO levels ($g=3\Delta$, $s=4\Delta$,
$\Omega=10\Delta$, $\kappa=0.014$, $k_BT=0.5\hbar\Delta$,
$\Omega_p=4\Delta$). Inset ($a$): NIBA result for $k_B
T=2\hbar\Delta$. The arrows indicates the first red sideband at
$\nu=\Omega+\Omega_p$. Inset ($b$): $Q'(t)$ vs $t$ shows damped
oscillations. } \label{niba}
\end{figure}

In the complementary regime of large environmental
coupling and/or high temperatures it is convenient to  employ
model ($a$),
and it is appropriate to treat the system's dynamics within the
noninteracting-blip approximation (NIBA) \cite{Wei}. The NIBA  is
non-perturbative in the coupling  $\alpha$
 but perturbative in the tunneling splitting $\Delta$. It is
  a good approximation for sufficiently high temperatures and/or
dissipative strength, or for symmetric systems.
Within the NIBA and in the limit of large driving frequencies
$\Omega\gg\Delta$, one finds \cite{Gri98}
\begin{equation}
P_{\infty}=\frac{k_0^-(0)}{k_0^+(0)}\,.
\end{equation}
Here,
\begin{eqnarray}
k^{-}_0(0)&=&\Delta^2 \int_0^\infty dt h^{-}(t)
\sin{(\varepsilon_0t)} J_{0}\left(\frac{2s}{\Omega}
\sin{\frac{\Omega t}{2}}\right)\, , \nonumber\\
k^{+}_0(0)&=&\Delta^2 \int_0^\infty dt
h^{+}(t)\cos{(\varepsilon_0 t)}
J_{0}\left(\frac{2s}{\Omega}\sin{\frac{\Omega t}{2}}\right)\;, \nonumber \\ 
\label{kernels}
\end{eqnarray}
%
with  $J_0$ being the zeroth order Bessel function. Dissipative
effects of the environment are
 captured by the terms
\begin{eqnarray}
h^{+}(t)&=&e^{-Q'(t)}\cos[Q''(t)]\,,\nonumber\\
h^{-}(t)&=&e^{-Q'(t)}\sin[Q''(t)]\,.
\end{eqnarray}
Here,  $Q'(t)$ and $Q''(t)$ are the real and imaginary parts of the bath
correlation function
\begin{eqnarray}
Q(t)=\int_0^\infty d\omega
\frac{J(\omega)}{\omega^2}\,
\frac{\cosh(\omega\beta/2)-\cosh[\omega(\beta/2-it)]}{\sinh(\omega\beta/2)}\,.
\nonumber
\end{eqnarray}
For the peaked  spectral density given in Eq.\  (\ref{jeff}) one finds
\begin{eqnarray}
Q'(t)&=& Q'_1(t)-e^{-\Gamma t}[Y_1\cos(\bar{\Omega}_{p}t)
+Y_2\sin(\bar{\Omega}_{p}t)]\, , \nonumber\\
 Q''(t)&=&A_1-e^{-\Gamma
t}[A_1\cos(\bar{\Omega}_pt)+A_2\sin(\bar{\Omega}_{p}t)] \, .
\label{qqprime}
\end{eqnarray}
Here,  $\beta=\hbar/k_B T, \Gamma=\pi\kappa\Omega_{p}$,
$\bar{\Omega}_{p}^2=\Omega_{p}^2-\Gamma^2$ and
\begin{eqnarray}
Q'_1(t)&=&
Y_1+\pi\alpha\Omega_p^2 \left[\frac{\sinh(\beta\bar{\Omega}_p)t}{2C\bar{\Omega}_p}+\frac{\sin(\beta\Gamma)t}{2C\Gamma} \right.
 \nonumber \\
&&  \left. -\frac{4\Omega_{p}^2}{\beta}\sum_{n=1}^\infty
\frac{\frac{1}{\nu_n}[e^{-\nu_n
t}-1]+t}{(\Omega_p^2+\nu_n^2)^2-4\Gamma^2\nu_n^2} \right],
\label{kern}
\end{eqnarray}
where $\nu_n=2\pi n/\beta$. Moreover,
$C=\cosh(\beta\bar{\Omega}_{p})-\cos(\beta\Gamma)$, $CY_{1/2}=\mp
A_{2/1}\sinh{(\beta\bar{\Omega}_{p})}-A_{1/2}\sin{(\beta\Gamma)}$,
$A_2=\alpha\pi(\Gamma^2-\bar{\Omega}_p^2)/(2\Gamma\bar{\Omega}_p)$,
$A_1=\pi\alpha$. As follows from Eq.\  (\ref{qqprime}),  $Q'$ and $Q''$
display damped oscillations with frequency $\bar{\Omega}_p$
(cf.\ Fig.\ \ref{niba}b) which are not present for
a pure Ohmic spectrum. It is the interplay between these
oscillations and the driving field which induces the extra
resonances in $P_\infty$.

To proceed, we rewrite the kernels $k_0^{\pm}(0)$ in a more
convenient form. In the integrand of Eq.\ (\ref{kernels}) the
functions $\cos[Q''(t)]$, $\sin[Q''(t)]$ and $e^{-Q'(t)+Q'_1(t)}$
oscillate with frequency $\bar{\Omega}_p$ and we can expand
them as
\begin{eqnarray}
\cos[Q''(t)]=\sum_{m=-\infty}^{\infty} \left[
D_m\cos(m\bar{\Omega}_pt)+E_m
\sin(m\bar{\Omega}_pt)\right] \, ,\nonumber \\
\sin[Q''(t)]=\sum_{m=-\infty}^{\infty}\left[ F_m\cos(m\bar{\Omega}_pt)+G_m
\sin(m\bar{\Omega}_pt)\right]\, ,\nonumber \\
e^{-Q'(t)+Q'_1(t)}=\sum_{m=-\infty}^{\infty}\left[
H_m\cos(m\bar{\Omega}_pt) +K_m\sin(m\bar{\Omega}_pt)\right]\, . \nonumber
\\
\label{fexpansion}
\end{eqnarray}
The coefficients $D_m$, $E_m$, $F_m$, $G_m$, $H_m$ and $K_m$ are
time dependent and they are given in Appendix \ref{app.niba}.
Inserting these expansions into Eq.\ (\ref{kernels}), and also
using the Fourier representation of
$J_0\left(\frac{2s}{\Omega}\sin{\frac{\Omega t}{2}}\right)$, we find
\begin{equation}
k_0^{\pm}(0)=\sum_{m=-\infty}^{\infty}\sum_{n=-\infty}^\infty\Delta^2\int_0^\infty
dt e^{-Q_1'(t)}f^{\pm}_{mn}(t)\,.
\label{kernelsimple}
\end{equation}
Here, $\varepsilon_{mn}=\varepsilon_0-m\bar\Omega_p-n\Omega$, and
\begin{eqnarray}
\label{expansion} &&f^{\pm}_{mn}(t)={{\rm Re} \atop {\rm
Im}}\left[{c^{\pm}_{mn}(t)\cos(\varepsilon_{mn}t)\pm c^{\mp}_{mn}(t)\sin(\varepsilon_{mn}t)}
\right],\nonumber\\
&&c^+_{mn}=J_n^2\left(\frac{s}{\Omega}\right)J_m(e^{-\Gamma t}\omega_1)
\cos(m\phi)(-i)^m e^{-iA_1}, \nonumber\\
&&c^-_{mn}=J_n^2\left(\frac{s}{\Omega}\right)J_m(e^{-\Gamma t}\omega_1)
\sin(m\phi)(-i)^m e^{-iA_1}\, ,
\end{eqnarray}
with $J_n$ being a Bessel function of order $n$, and
\begin{eqnarray}
\omega_1&=&\sqrt{(A_1-iY_1)^2+(A_2-iY_2)^2}\,, \nonumber\\
\tan\phi&=&-\frac{A_2-iY_2}{A_1-iY_1}\, .
\end{eqnarray}
Thus, from Eq.\ (\ref{expansion}),
 we expect resonances when $\varepsilon_{nm}=0$.

In the limit $\Gamma/\Omega_p\ll 1$
 and for not too large $T$ (i.e.,  $\cos(\beta\Gamma)\ll\cosh(\beta\Omega_p)$),
 we find that
 \begin{eqnarray}
\tan(m\phi)\approx i\tanh\left(\frac{m\beta\Omega_p}{2}\right)\,.
\end{eqnarray}
Inserting this into Eq.\ (\ref{expansion}), we obtain
\begin{eqnarray}
i\tanh\left(\frac{m\beta\Omega_p}{2}\right)c_{mn}^+=c_{mn}^-\,.
\end{eqnarray}
If the environmental mode is enough localized (i.e., the integrand
of Eq.\ (\ref{kernels}) is only damped after several
oscillations), we expect that the sum in Eq.\ (\ref{kernelsimple})
is dominated by the coefficient of $\cos{\varepsilon_{nm}t}$ if
$\varepsilon_{mn}=0$. This means that
\begin{eqnarray}
f_{mn}^+(t)&\approx &{\rm Re}[c_{mn}^+(t)]\,\nonumber\\
f_{mn}^-(t)&\approx&\tanh\left(\frac{m\beta\Omega_p}{2}\right){\rm
Re}[c_{mn}^+(t)]\, ,
\end{eqnarray}
which leads to
\begin{eqnarray}
P_{\infty}=\tanh\left(\frac{m\beta\Omega_p}{2}\right)\, .
\label{Pinftyniba}
\end{eqnarray}
Without driving we only have terms with $n=0$ and $\varepsilon_{m0}=0$
implies that $\varepsilon_0=m\Omega_p$.
In that case Eq.\ (\ref{Pinftyniba}) gives the NIBA thermal equilibrium value.
Hence, in order to find resonances we need to apply driving.
 For ``conventional'' resonances at $\varepsilon_0=n\Omega$,  we put $m$ to
 zero and we find  $P_\infty\approx 0$, as predicted
for unstructured  environments \cite{Har00,Goo03}. Finally, for
$\varepsilon_0=n\Omega\pm m\Omega_{p}$, we recover
$P_\infty\approx\pm\tanh(m \beta\Omega_{p}/2)$, as was also found
within the Floquet-Born-Markov approach, cf.\  (\ref{pinffloquet}).
 Results of a numerical evaluation of
$P_\infty$ are shown in Fig.\  \ref{niba}, using the NIBA result
(\ref{expansion}), as well as the exact ab-initio real-time
 QUAPI method \cite{QUAPI,Tho98,Tho00}. 
Resonance dips are
observed at $\varepsilon_0=\Omega$,
$\varepsilon_0=\Omega-\Omega_p$ and
$\varepsilon_0=\Omega-2\Omega_p$. For $k_B T\sim \hbar \Omega_p$,
we also find the first red sideband at
$\varepsilon_0=\Omega+\Omega_p$, see inset (a).

\section{Limit $\Omega_p \gg \nu $ \label{sec.bessel}}

In the limit when the frequency $\Omega_p$ of the HO is much
larger than the effective TSS level splitting $\nu$, the peak in
the spectral density at $\Omega_p$ acts as a high-frequency
cut-off for an effective Ohmic bath, see Eq.\  (\ref{jeff}). In
other words, the oscillations in the correlation functions, see
Sec.\  \ref{sec.strongdamp} which occur on a time-scale
$\Omega_p^{-1}$ are very fast and can be averaged out when only
the long-time dynamics is of interest and short-time effects are
not considered. In this limit, the standard driven and Ohmically
damped spin-boson model \cite{Wei,Gri98} is recovered.  In the
regime of weak damping $\alpha \ll 1$, the stationary population
difference $P_\infty$ has been determined within the assumption of
large driving frequencies $\Omega\gg \Delta,...$ upon using
analytic real-time path integral methods in Ref.\
\onlinecite{Har00}. In this Section, we use this high-frequency
approximation of Ref.\ \onlinecite{Har00} as a starting point, and
derive a closed simple analytic expression for the peak shape of
the ``common'' multi-photon resonance. Most importantly, we find
the scaling of the width of the $n$-photon resonance as the $n$-th
Bessel function $J_n(s/\Omega)$. This scaling behavior has been
observed experimentally in superconducting flux qubit devices
\cite{Saito04}.

The central issue in finding a closed analytic expression for $P_\infty$ is to
find the roots $\vartheta_n$ of the pole equation \cite{Har00}
\begin{equation}
\prod_{n=-\infty}^{+\infty} \left( \epsilon_n^2 -\vartheta^2 \right) +
\sum_{n=0}^{+\infty} \Delta_n^2 \prod_{m=-\infty, m\ne n}^{+\infty}
 \left( \epsilon_m^2 -\vartheta^2 \right) = 0 \, .
\label{poleeq}
\end{equation}
Here, $\epsilon_n=\varepsilon_0-n\Omega$ is the photon-induced
bias and $\Delta_n=|J_n(s/\Omega)| \Delta$ is the field-dressed
tunneling splitting of the TSS, where $J_n(x)$ is the $n$-th
ordinary Bessel function. Considering the $n$-photon resonance, we
 numerically find that,  up to
extremely high numerical precision, the roots
 of Eq.\  (\ref{poleeq}) are given by
\begin{eqnarray}
\vartheta_0& = & \sqrt{\varepsilon_0^2+\Delta_0^2} \, ,\nonumber \\
\vartheta_{n\ne0} & = & \sqrt{\epsilon_n^2+\Delta_n^2} \, , \nonumber \\
\vartheta_{k\ne n,0} & = &  \epsilon_k \, .
\label{roots}
\end{eqnarray}
Plugging  Eqs.\ (\ref{roots}) in the expressions for $P_\infty$
given in Ref.\ \onlinecite{Har00}, see Eqs.\  (6) and (7) therein,
we find a closed expression for the lineshape of the $n$-photon
resonance to be 
\begin{widetext}
\begin{equation}
P_\infty^{(n)}(\varepsilon_0) =
\frac{
\varepsilon_0 \epsilon_n^2\Delta_0^2 (\varepsilon_0^2+\Delta_0^2-\epsilon_n^2)
\frac{J_{\rm Ohm}(\vartheta_0)}{\vartheta_0}
+
\varepsilon_0^2 |\epsilon_n|\Delta_n^2 (\epsilon_n^2+\Delta_n^2-\varepsilon_0^2)
\frac{J_{\rm Ohm}(\vartheta_n)}{\vartheta_n}
}
{
\Delta_0^2\epsilon_n^2(\varepsilon_0^2+\Delta_0^2-\epsilon_n^2)
J_{\rm Ohm}(\vartheta_0)   \coth \frac{\hbar\vartheta_0}{2k_B T}
-
\Delta_n^2\varepsilon_0^2(\epsilon_n^2+\Delta_n^2-\varepsilon_0^2)
J_{\rm Ohm}(\vartheta_n)   \coth \frac{\hbar\vartheta_n}{2k_B T}
} \, .
\label{multiphotonpeak}
\end{equation}
\end{widetext}
This result can be simplified  upon observing that the second term in the
numerator is small if the driving is not too large, since then $\Delta_n^2 \ll
\Delta_0^2$. Moreover, we are interested in the regime $\epsilon_0 \gg \Delta$
which is the saturation regime implying that $\nu\approx \varepsilon_0$ and at
low temperatures. Then,
away from the resonance point at $n \Omega \approx \varepsilon_0$, the second
term in the denominator in Eq.\  (\ref{multiphotonpeak}) can be neglected and
one recovers the standard result, if one uses that $\Delta_0\approx \Delta$
which is fulfilled for weak driving. It reads
\begin{equation}
P_\infty^{(n)}(\varepsilon_0) =
\frac{\varepsilon_0}{\sqrt{\varepsilon_0^2+\Delta^2}}
\tanh \frac{\sqrt{\varepsilon_0^2+\Delta^2}}{2k_B T} \, ,
\end{equation}
which gives the correct result away from any $n$-photon resonance.
For the case at the $n$-photon resonance at $n \Omega \approx \varepsilon_0$,
one finds a Lorentzian line shape, i.e.,
\begin{equation}
P_\infty^{(n)}(\varepsilon_0) =
\frac{\Delta_0^2(\varepsilon_0-n\Omega)^2}
{\Delta_0^2(\varepsilon_0-n\Omega)^2+2\varepsilon_0^2\Delta_n^2k_B T / \hbar}
\, ,
\label{respeak}
\end{equation}
where we have expanded the second $\coth$ term in the denominator
(see Eq.\ (\ref{multiphotonpeak})) up to lowest order in the
argument, which is appropriate since $\vartheta_n$ is small at
resonance (and at low temperature). The linewidth of the
Lorentzian peak can be calculated as the full width at half
maximum (FWHM)
\begin{equation}
\Delta \varepsilon^{(n)} = 2 \sqrt{2 n \Omega \left(
\frac{\Delta_n}{\Delta_0}\right)^2 \frac{k_B T}{\hbar} +
\left(\frac{\Delta_n}{\Delta_0}\right)^4 \left(\frac{k_B T}{\hbar} \right)^2} \,
.
\label{fwhm1}
\end{equation}
Note that this result obtained from the high-frequency approximation is
independent of the damping constant. Moreover, the leading term is the first
term under the square root in Eq.\  (\ref{fwhm1}).
Note furthermore  that for the case of infinitesimal driving, the FWHM is not
correctly reproduced by Eq.\  (\ref{fwhm1}) since it would approach zero.
However, as it is known from NMR within a treatment in terms of the Bloch
equation, in this case, the FWHM is dominated by the dephasing \cite{Goo03}, i.e.,
\begin{equation}
\Delta \varepsilon^{(1)}_{\rm Bloch} = 2 \sqrt{\Gamma_{\phi}^2 + \Omega_R^2
\Gamma_\phi/\Gamma_R} \, ,
\end{equation}
where $\Gamma_{R}=\pi \alpha \coth (\hbar \nu/2k_B T) \Delta^2/\nu$
is the relaxation rate
and $\Gamma_\phi= \Gamma_R/2 + 2 \pi \alpha (\varepsilon_0^2/\nu^2) k_BT/\hbar$
\cite{Wei}.
Both rates are of first order in the damping strength $\alpha$. Moreover,
 $\Omega_R$
is the (single-photon) Rabi frequency. Hence, we have to include the dephasing rate
$\Gamma_\phi^2$ in Eq.\  (\ref{respeak}), since it cannot be reproduced by our weak-coupling
approach which is only of first order in $\alpha$. This finally yields in
leading order in the driving strength
\begin{equation}
\Delta \varepsilon^{(n)} = 2 \sqrt{\Gamma_\phi^2 +
\left(\frac{\Delta_n}{\Delta_0}\right)^2 2 n \Omega
 \frac{k_B T}{\hbar}} \, .
\label{fwhm2}
\end{equation}
As follows from Eq.\  (\ref{fwhm2}), the FWHM  of the $n$-photon resonance
scales with the $n$-th ordinary Bessel function, i.e.,
$\Delta \varepsilon^{(n)} \sim J_n(s/\Omega)$ as also confirmed by experimental
measurements \cite{Saito04}.

\section{Conclusions\label{sec.conclusio}}

In conclusion we have investigated the problem of a quantum
mechanical driven two-state system being coupled to a structured
environment which has a localized mode at a frequency $\Omega_p$
but behaves Ohmically at low frequencies. We have studied two
complementary parameter regimes of weak and strong coupling to
the environment. The interplay of the driving and the localized
mode gives additional features like resonant peaks/dips in the
asymptotic averaged population difference $P_\infty$. We have
calculated analytically the lineshape of the resonances in various
parameter regimes and have obtained simple closed expressions for
the particular example of the first blue sideband. We also include
the discussion of how
 the results are generalized for any sideband. Moreover, we have
elaborated the limit when the localized mode acts as a high-frequency cut-off.
Then, the full width at half maximum of the $n$-photon resonance
has been shown to scale with the $n$-th ordinary Bessel function.


Our model finds as well applications in the field of
cavity quantum electrodynamics (CQED) with solid state structures
\cite{Wallraffreview}.  Most interestingly, the strong coupling limit of CQED
could be reached in superconducting electrical circuits, with perspective
applications ahead.

Finally, we note that a related experiment has  been reported
recently by Wallraff {\em et al.\/} \cite{Wallraff04}. There, a
qubit was realized in the form of a Cooper pair box which couples
to a single mode of a cavity which is damped. The properties of
the TSS-HO were probed spectroscopically by measuring the
transmission of the resonator. In other words, a {\em driven\/} HO
was considered while the TSS was kept static. In contrast to that
system, here, the TSS was time-dependent while the HO is treated
as static.

\begin{acknowledgments}
We thank  P. Bertet, I. Chiorescu and H. Mooij for discussions.
This work has been supported by the Universit\"atsstiftung Hans Vielberth and the Dutch NWO/FOM. \\
\end{acknowledgments}

\appendix

\section{Symmetry properties for the dissipative rates for the first blue
sideband \label{app.sym}}
In order to evaluate the stationary averaged population difference
$P_\infty$, the rate coefficients $L_{\alpha\beta,\alpha^\prime
\beta^\prime}$ have to be determined explicitly. For the example of the
resonance at $\nu\approx\Omega-\Omega_{p}$ (first blue sideband)
considered in this work, we find that the rate coefficients fulfill the
symmetry relations
\begin{widetext}
\begin{eqnarray}
&&L_{11,22}=L_{22,11}=L_{33,44}=L_{44,33}=0,\,\,\,\,\,
L_{11,11}=-(L_{33,11}+L_{44,11}), \nonumber\\
&&L_{22,22}=-(L_{33,22}+L_{44,22}),\,\,\,\,\,
L_{33,33}=-(L_{11,33}+L_{44,33}),\nonumber\\
&&L_{44,44}=-(L_{11,44}+L_{22,44}),\,\,\,\,\,
L_{12,12}=(L_{22,22}+L_{11,11})/2,\,\,\,\,\,
L_{11,33}=L_{44,22},\nonumber\\
&&L_{22,33}=L_{44,11},\,\,\,\,\,
L_{33,22}=L_{11,44},\,\,\,\,\,
L_{33,11}=L_{22,44},\,\,\,\,\,
L_{22,21}=L_{21,11},\nonumber\\
&&L_{11,21}=L_{21,22},\,\,\,\,\,
L_{44,21}=-L_{21,33},\,\,\,\,\,
L_{33,21}=-L_{21,44},\,\,\,\,\,
L_{22,12}=L_{22,21},\nonumber\\
&&L_{11,12}=L_{11,21},\,\,\,\,\,
L_{44,12}=L_{44,21},\,\,\,\,\,
L_{33,21}=L_{33,12},\,\,\,\,\,
L_{12,21}=0.
\label{symmetries}
\end{eqnarray}
\end{widetext}
As a consequence, there are eight independent rates given by
\begin{widetext}
\begin{eqnarray}
L_{44,22}& = & 2L_{\rm
osc}(\varepsilon_{4,2,-1})\cos^2\left(\frac{\theta}{2}\right)
+2L_{\rm q}(\varepsilon_{4,2,0})\sin^2\left(\frac{\theta}{2}\right)\nonumber-L_{\rm q,osc}
(\varepsilon_{4,2,-1})\sin \theta \, , \nonumber\\
L_{22,44}&=&2L_{\rm osc}(\varepsilon_{2,4,1})\cos^2\left(\frac{\theta}{2}\right)
+2L_{\rm q}(\varepsilon_{2,4,0})\sin^2\left(\frac{\theta}{2}\right)\nonumber-L_{\rm q,osc}
(\varepsilon_{2,4,1})\sin\theta \, , \nonumber\\
L_{44,11}&=&2L_{\rm osc}(\varepsilon_{4,1,-1})\sin^2\left(\frac{\theta}{2}\right)
+2L_{\rm q}(\varepsilon_{4,1,0})\cos^2\left(\frac{\theta}{2}\right)\nonumber
+L_{\rm q,osc}(\varepsilon_{4,1,-1})\sin\theta \, , \nonumber\\
L_{11,44}&=&2L_{\rm osc}(\varepsilon_{1,4,1})\sin^2\left(\frac{\theta}{2}\right)
+2L_{\rm q}(\varepsilon_{1,4,0})\cos^2\left(\frac{\theta}{2}\right)\nonumber
+L_{\rm q,osc}(\varepsilon_{1,4,1})\sin\theta \, , \nonumber\\
L_{21,22}&=&\frac{1}{2}\left(L_{\rm osc}(\varepsilon_{3,2,0})
-L_{\rm q}(\varepsilon_{3,2,-1})-L_{\rm osc}(\varepsilon_{4,2,-1})
+L_{\rm q}(\varepsilon_{4,2,0})\right)\sin\theta+\frac{1}{2}(L_{\rm q,osc}(\varepsilon_{3,2,0})
-L_{\rm q,osc}(\varepsilon_{4,2,-1}))\cos\theta \, , \nonumber\\
L_{21,11}&=&\frac{1}{2}\left(L_{\rm osc}(\varepsilon_{3,1,0})-L_{\rm q}(\varepsilon_{3,1,-1})
-L_{\rm osc}(\varepsilon_{4,1,-1})+L_{\rm q}(\varepsilon_{4,1,0})\right)\sin\theta
+\frac{1}{2}(L_{\rm q,osc}(\varepsilon_{3,1,0})
-L_{\rm q,osc}(\varepsilon_{4,1,-1}))\cos\theta \, , \nonumber\\
L_{21,44}&=&\frac{1}{2}\left(L_{\rm osc}(\varepsilon_{1,4,1})
+L_{\rm osc}(\varepsilon_{2,4,1})-L_{\rm q}(\varepsilon_{1,4,0})
-L_{\rm q}(\varepsilon_{2,4,0})\right)\sin\theta+\frac{1}{2}(L_{\rm q,osc}(\varepsilon_{1,4,1})
+L_{\rm q,osc}(\varepsilon_{2,4,1}))\cos\theta \, , \nonumber\\
L_{21,33}&=&\frac{1}{2}\left(L_{\rm q}(\varepsilon_{1,3,1})+L_{\rm q}(\varepsilon_{2,3,1})
-L_{\rm osc}(\varepsilon_{1,3,0})-L_{\rm osc}(\varepsilon_{2,3,0})\right)\sin\theta\nonumber
-\frac{1}{2}(L_{\rm q,osc}(\varepsilon_{1,3,0})
+L_{\rm q,osc}(\varepsilon_{2,3,0}))\cos\theta \, ,
\label{allrates}
\end{eqnarray}
with
\begin{eqnarray}
L_{\rm q}(\varepsilon_{klm})&=&\langle e,1,0|e^{iS}Xe^{-iS}|g,1,0\rangle^2N(\varepsilon_{klm})
=\langle e,0,0|e^{iS}Xe^{-iS}|g,0,0\rangle^2N(\varepsilon_{klm})
=\frac{4g^2\Delta^2\Omega_p^2N(\varepsilon_{klm})}
{\nu^2(\nu^2-\Omega_p^2)^2} \, , \nonumber\\
L_{\rm osc}(\varepsilon_{klm})&=&\langle g,1,0|e^{iS}Xe^{-iS}|g,0,0\rangle^2N
(\varepsilon_{klm})=\langle e,1,0|e^{iS}Xe^{-iS}|e,0,0\rangle^2N
(\varepsilon_{klm})\nonumber\\
&=&\left(\frac{4g^2(\Delta^2(\nu^2-2\Omega_p^2)
-(\nu^2-\Omega_p^2)^2)}{\Omega_p^2(\nu^2-\Omega_p^2)^2}
+1\right)N(\varepsilon_{klm}) \, , \nonumber\\
L_{\rm q,osc}(\varepsilon_{klm})&=&-2\langle e,1,-1|e^{iS}
Xe^{-iS}|g,1,0\rangle\langle g,1,0|e^{iS}Xe^{-iS}|g,0,0\rangle
N(\varepsilon_{klm})\nonumber\\
&=&2\langle g,0,1|e^{iS}Xe^{-iS}
|e,0,0\rangle\langle e,0,0|e^{iS}Xe^{-iS}|e,1,0\rangle N(\varepsilon_{klm})\nonumber\\
&=&\frac{\Delta\varepsilon_0 gs((\Omega+\Omega_p)^2
+2\Omega_p^2+\nu(-\Omega+\Omega_p))N(\varepsilon_{klm})}
{2\nu(\nu-\Omega)\Omega(\nu-\Omega-\Omega_p)\Omega_p(\nu+\Omega_p)} \, , \nonumber\\
N(\varepsilon_{klm})&=&N(\varepsilon_k-\varepsilon_l+m\Omega) \, .
\label{transitions}
\end{eqnarray}
\end{widetext}

Note that $L_{\rm osc}$ is the rate containing the zeroth order
term in $g$ and $s$. It is related to the transition between two
states differing by one oscillator quantum. This decay is of
zeroth order (hence fast) because the oscillator is coupled
directly to the environment. Moreover, $L_{\rm q}$ gives the rate
for transitions between the excited and ground state of the TSS
with the HO remaining in the same state, and $L_{\rm q,osc}$ is
related to the transition where both the qubit and the oscillator
exchange energy with the environment. Note that this transition is
induced by the driving and involves one photon.

\section{Coefficients for the kernels $k_0^{\pm}(0)$ \label{app.niba}}
In Sec.\  \ref{sec.strongdamp}, we have introduced an
expansion of the oscillating functions given in Eq.\
(\ref{fexpansion}). In this appendix we summarize the
corresponding coefficients for completeness.

For the expansion of $\cos[Q''(t)]$,  we find
\begin{eqnarray}
D_{2m+1}&=&(-1)^m\sin{(A_1)} J_{2m+1}(A)\cos{[(2m+1)X]}\, ,\nonumber\\
D_{2m}&=&(-1)^m\cos{(A_1)} J_{2m}(A)\cos{(2mX)}\, ,\nonumber\\
E_{2m+1}&=&(-1)^m\sin{(A_1)} J_{2m+1}(A)\sin{[(2m+1)X]}\, ,\nonumber\\
E_{2m}&=&(-1)^m\cos{(A_1)} J_{2m}(A)\sin{(2mX)},
\end{eqnarray}
where we have introduced
\begin{eqnarray}
A&=&e^{-\Gamma t} \sqrt{A_1^2+A_2^2}\, ,\nonumber\\
\sin{X}&=&A_2/\sqrt{A_1^2+A_2^2}\, .
\end{eqnarray}
In the same way, the expansion of $\sin[Q''(t)]$ gives
\begin{eqnarray}
F_{2m+1}&=&(-1)^{m+1}\cos{(A_1)} J_{2m+1}(A)\cos{[(2m+1)X]}\, ,\nonumber\\
F_{2m}&=&(-1)^m\sin{(A_1)} J_{2m}(A)\cos{[2mX]},\nonumber\\
G_{2m+1}&=&(-1)^{m+1}\cos{(A_1)} J_{2m+1}(A)\sin{([2m+1)X]}\, ,\nonumber\\
G_{2m}&=&(-1)^m\sin{(A_1)} J_{2m}(A)\sin{[2mX]}\, .
\end{eqnarray}
Finally, we find for the coefficients of
 $\exp[Q(t)-Q_1(t)]$
\begin{eqnarray}
H_{m}&=&I_{m}(Y)\cos{(mV)}\,,\nonumber\\
K_m&=&I_{m}(Y)\sin{(mV)}\,,
\end{eqnarray}
where we have introduced
\begin{eqnarray}
Y&=&e^{-\Gamma t} \sqrt{Y_1^2+Y_2^2}\, ,\nonumber\\
\tan{V}&=&\frac{Y_2}{Y_1}\, .
\end{eqnarray}
Here, $I_m$ is the modified Bessel function of the first kind of
order $m$.

\end{document}